# Femtosecond Laser Engraved 2D Tunable Optofluidic Liquid Core/Air Cladding Channel Waveguides on PDMS

Sanyogita*, Amar Ghar and P. K. Panigrahi

Centre for Lasers and Photonics, Indian Institute of Technology, Kanpur-208016 (UP).

sanyogita.iitk@gmail.com

We have demonstrated fabrication and characterization of 2D liquid based multimode optical waveguide structures over Polydimethylsiloxane (PDMS) material based chip. Fabrication of two separate microsturures, one with width of 14 micron and depth of 27 micron while the other with width as well as depth of 110 micron, was achieved by femtosecond laser micromachining process. The dye solution is passed through the microstructure from one end to the other; wherein dye solution acts as the core while PDMS and air act as cladding medium. The femtosecond laser micromachining parameters are optimized in terms of laser power, pulse width, writing speed, focused beam size etc. Quality of fabricated microstructures is confirmed by microscopic analysis. The confirmation of liquid core/air cladding based waveguide is obtained through the spectral and modal analysis. The optical analysis has been done by using fluorescence light coupled out from waveguide structures filled with different dye solutions. These waveguide structures give strong light confinement and intense interaction between dye solution and pump light. The developed micro structures are tunable in terms of intensity, wavelength and beam size. Such micro structures can be implemented in design and development of lab-on-chip micro lasers and sensing applications in any multifunction lab-on-chip devices.

## Introduction

Optofluidic is a great research platform where the advantages of both optics and microfluidics can be combined in a single chip to move towards highly compact, portable and multifunctional devices [1]. This optofluidic lab-on-a-chip (LOC) approach provides a huge potential in terms of low-cost optical sources, sensors, liquid-liquid waveguide, liquid core waveguide and real time detection. Particularly in photonic science, and more specifically in the micro and nano regime, the integration of fluid and light in the same path offers the capacity to reconfigure the device in accordance with the choice of fluid opted as the fluid medium and thus providing dynamic and powerful practical tuning mechanism, making it customizable in real time [2, 3].

Nonetheless, the fabrication and characterization process are complicated owing to the miniscule dimensions of such microstructures and managing the required smoothness at the edges of microchannel and waveguide wall. High precision handling of chip is also a must to minimize optical losses and for accurate control over light and fluid in the micro/nano regime to maintain good functionality. In the liquid core/air cladding waveguide chip, the refractive index of core material has to be higher than that of the cladding so as to enable total internal reflection (TIR) phenomenon for the refractive index guided mode. Moreover, dye solutions with different host materials and concentrations have broad range variation in refractive index to that of water. Such an enhanced range helps in sustaining the liquid core-air waveguide over the long flow path for a much higher operational time. This feature provides for a substantial increase in wider applications of mode for such type of optofluidic chip.

Optofluidic waveguides can confine light in small dimensions and generate high intensity optical beam over a long distance, creating a potential for tremendous applications in the field of environmental monitoring, bio-sensing, analytical chemistry etc. [4].

Various methods have been proposed to fabricate 2D structures; among them, structure fabrication using soft lithography process is widely prevalent [5,6]. But the soft lithography process in itself have a number of disadvantages like involvement of multiple fabrication steps, high rate of errors while achieving required depth of microstructures, longer time of fabrication etc. Most noticeable drawback of soft lithography is that it requires another lithography method such as photolithography or e-beam lithography to fabricate the stamp master used in further development process of microstructure [6]. On the other hand, Femtosecond laser based direct writing has many advantages over other conventional methods such Excimer laser writing, $CO_2$ laser writing-beam lithography and soft lithography etc.[6,7] for fabrication of microstructures. Femtosecond laser interaction with soft materials has opened up a new field of waveguide fabrication methods for structures on the surface as well as inside of transparent materials. A femtosecond laser emits pulsed beams with durations of tens or hundreds of femtosecond region which, nowadays, are used for high-quality micro and nanofabrication. As the energy deposition time of femtosecond laser is shorter than time required to release the energy in the form of heat using electron-photon coupling process, heat affected zone is completely suppressed during the laser pulse interaction even with soft material like PDMS [7]. This feature enables laser processing on PDMS with high precision and resolution. Another advantage of femtosecond laser processing over conventional methods is the capability of sculpturing complex shapes at micro and nanoscale in transparent materials. With the help of focused fs-laser beam one can achieve extremely high peak intensity in the focused region which provides for high precision in setting up interaction region at the surface or even inside the volume. This feature not only eliminates a complicated and multiple patterning processing, involved in the conventional methods like photolithography for 2D fabrication, but also makes it feasible to create complex 2D structures which were not easily achievable by other conventional methods. The application of femtosecond micromachining to develop the optofluidic devices improves their structural and optical qualities to such an extent that it could provide a major alternate platform to innovate and produce novel optical devices on mass production level. Hence, this unique technique is going to contribute as a promising tool in the photonics fields and will help in emergence of new businesses once it reaches commercialization.

In this paper, we have demonstrated the fabrication of micro structures by using femtosecond direct writing along with development of liquid core-based waveguide. Structuring of 2D micro channels on the surface of PDMS is fabricated by f-s laser. These microchannels are converted to a super hydrophobic nature which can provide for an effective wave guiding. For light flow path, R6G and RH101 dye solutions were selected as liquid core medium. These dyes are distributed evenly along the length of the two prototypes that we have fabricated as two microchannels. Concentration of dye solution is chosen in such a way that refractive index of liquid medium is slightly higher than that of PDMS and air so that the PDMS and air ends up acting as a clad. Cross sections of these waveguide systems were captured by a CCD camera. Role of incident power, concentration of liquid dye and photo bleaching have been successfully studied thereof.

## Experimental Details

Femtosecond laser micromachining process has been used to fabricate two distinct dimensioned microstructures, each on a separate PDMS surfaces with a provision of inlet and outlet at the terminal ends for flow of liquid across the microchannel. These microchannel act as two unique liquid core/air clad waveguides. Fig. 1 shows the schematics of experimental set up for femtosecond laser-based micromachining system. The proposed experiment consists of regenerative Ti: Sapphire based amplified laser system (CLRK-MXR, USA) capable of delivering a maximum output power of 800 mW with pulse width of 120 fs having central wavelength of 775 nm and repetition rate of 1 KHz.

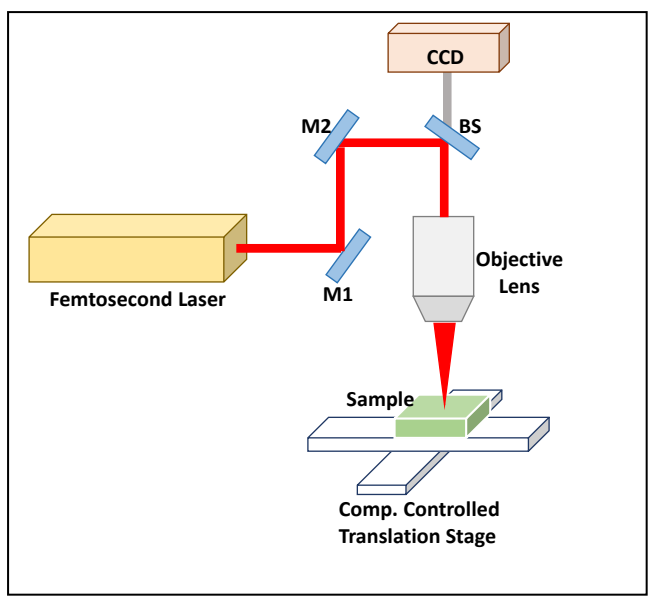

Fig. 1: Femtosecond micromachining fabrication setup for 2D Microstructures/hallow waveguide structure on PDMS

The output beam from fs-laser system is focused on surface of PDMS sample using 10X objective lens and beam aligning system (OPTEC Belgium). All the microstructures are created by successive translator movements of PDMS sample mounted on micro-position stage without any movement of focused laser beam. The PDMS substrate is irradiated with focused laser beam. The key steps in the experiment includes focusing lens and micro-position translation stage with 1 um resolution as shown in Fig 1. The focusing objective lenses are used to converge the laser beam providing a greater depth of field and smaller spot size as per the calculated requirement which is important for precision laser micro-machining process. Micro-position stage is used to move the sample as per the designed program. The computer-controlled laser power and micromachining system ensures that position errors and beam distortions are minimized over the entire scan region.

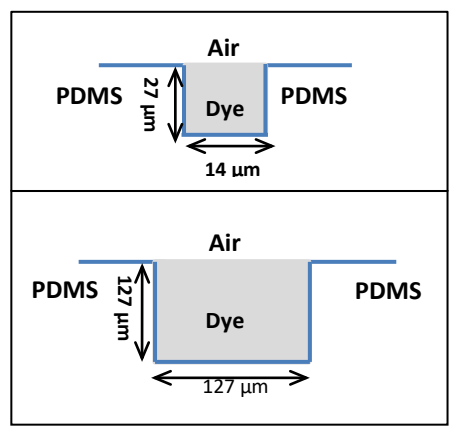

Fig. 2: Schematics of: a. Waveguide-I cross section; b. Waveguide-II cross section

For this experimental study, two straight microchannels on separate surfaces of PDMS have been fabricated successfully with different focusing lens. Both the microchannels are fabricated with different lasing power and focusing lenses. First microstructure (larger microchannel) is fabricated with a width of 110 μm and a depth of 110 μm and the second microstructure (smaller microchannel) with a width of 14 um and a depth of 27.937 μm as shown in Fig. 2. The larger microchannel has been fabricated by setting the laser power at 25 mW with a spot size of 15 μm (writing speed was kept 1mm/sec) and using multi-pass laser scan over the square shaped cross section. Based on multimode waveguide, the target cross-section is scanned 10 times horizontally and 5 times vertically with a beam overlap of 10 μm. Fabrication of inlet and outlet has also been done by fs-laser using multi-pass laser scan. The smaller microchannel (waveguide I) as well, has been fabricated with multi-pass laser scan but with slightly different writing parameters. Here laser power was taken as 18 mW with a beam spot size of 8 μm and horizontal scanning was done only twice with a beam overlap of 6 μm (writing speed 1mm/sec). After the measurement width of channel was found to be 14 μm and depth was 27.937 μm. In order to flow the dye solutions through fabricated channels, uniform inlet and outlet connected to central microchannels have also been fabricated with a multi-pass and multi scan using fs-laser. Inlet as well as outlet for bigger microchannels measure 110 μm in width and 40 μm in depth and for smaller microchannel width was 110 μm and depth was 20 microns. In both the cases we have kept the depth of inlet and outlet less than the central microchannel, for easy flow of liquid in to it from.

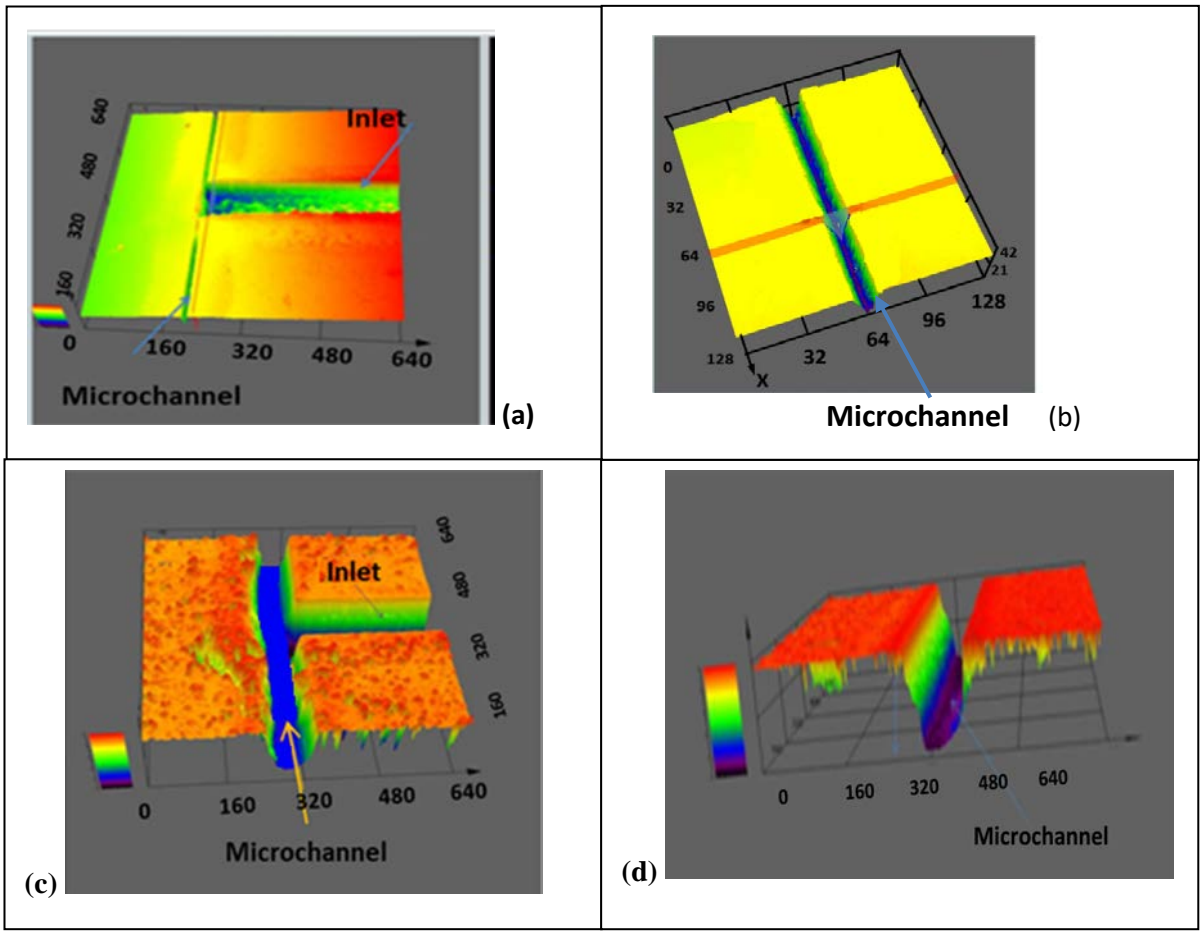

Fig. 3: (a) 2D waveguide structure-I over PDMS, (b) Cross section of Waveguide structure-I (c) 2D microstructure-II over PDMS (D) Cross section of microstructure-II

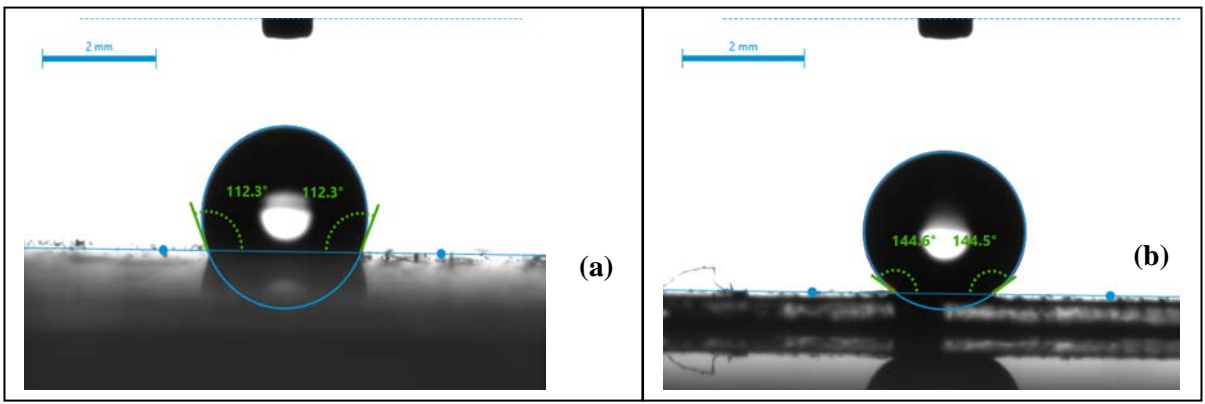

Fig. 4: Contact angle measurement for (a) Plane PDMS surface and (b) for PDMS surface exposed with femtosecond laser

The corresponding width and depth of developed microstructures have been confirmed by image analysis obtained with confocal microscope (Olympus LEXT OLS 4000) as shown in Fig. 3 above. This system capable of resolution up to 10 nm in Z direction and 120 nm in X-Y plane. The super hydrophobic channels are effective in creating air cladding between the dye filled liquid core and solid walls of PDMS, thus providing a good coupling for TIR and the waveguide. Here, due to 2D wave guiding, scattering and diffraction of visible light still persists to the channel walls. Light undergoes TIR at the front end of the channel too. Due to femtosecond structuring on the PDMS material, the PDMS channel wall is also made hydrophobic which controls the losses of waveguide. After measuring the contact angle for femtosecond direct-written 2D microchannel as shown in Fig. 4, the hydrophobicity was checked for the contact surface modified due to exposure of femtosecond laser with similar parameters that one used to fabricate microstructures on PDMS respectively. It was found that channel has been converted into a hydrophobic channel. These hydrophobic channels have low solid fraction that can effectively support the liquid-core/air cladding waveguide configuration on lab-on-chip platform. Hence, this unique structure allows an effective control and flow of light from one end to other.

## Implementation of microstructure as an optical waveguide

The two fabricated microchannels, with 2D square and rectangle shape cross section respectively are filled with liquid dye medium in order to convert it into liquid based multimode waveguide microstructures. The structures act as liquid-core waveguide platform when the refractive index (*n*) of cladding material (PDMS/air) is smaller than that of the flowing dye solution which acts as the core and enable the total internal reflection for the configuration of the index-guided mode [8, 9]

The waveguide losses are also sensitive to the roughness of the surfaces of the waveguide walls. As the waveguide walls are pretty smooth in case of femtosecond fabrication, the losses are very much minimized in comparison to other conventional fabrication methods. Other challenges and issues in these experiments are also resolved as gas (i.e., air) is used as cladding material [9, 10]. Air has a much lower refractive index ($n_{air}$=1.0) than most of the solid and liquid materials, thus it allows a wider range of incident angles. Air also has much lower viscosity than that of any liquid so that it can significantly reduce the hydrodynamic friction and Joule heating at the interface between the core and the cladding [10]. Higher refractive index difference between the liquid core and air cladding (Δn= 0.407) helps to increase the amount of light trapped inside the core and avoids the diffusional mixing problem normally observed in liquid to liquid L2 waveguide.

In presented case, two types of dyes have been used as the gain material to demonstrate the concept of liquid-air waveguide on a chip. First dye is Rhodamine-6G dissolved in ethanol and benzyl alcohol while the second one is Rhodamine-101 dissolved in mixture of ethanol + benzyl alcohol in a concentration range of 1mM to 5mM for both liquid core solutions. The corresponding change of refractive index of fluid observed by varying the dye solution concentration for both dye solutions is measured by the refractometer (Abbemat 500). The refractive index difference of core and clad has been selected between $10^{-3}$ to $10^{-2}$ for index for varying concentration form of R6G and Rh101 from 1% to 10 %. From measurement, it is evident that dye solutions with different concentration can act as two different liquid core medium with varying characteristics. For example, in case of 1mM concentration Rh-6G dye solution ($n_2$=1.4030) in mix solution of (ethanol + benzyl alcohol) is higher than that of cladding material i.e. air ($n_1$= 1) and PDMS ($n_3$= 1.40). Liquid filled channel acts as a core in this case wherein light propagates through liquid core waveguide by satisfying condition of total internal reflection. This has been demonstrated through the resulting fluorescence emerging at the other end of the waveguide. Characteristics are found to be drastically different between the gain materials as they are confined to the liquid-air interface.

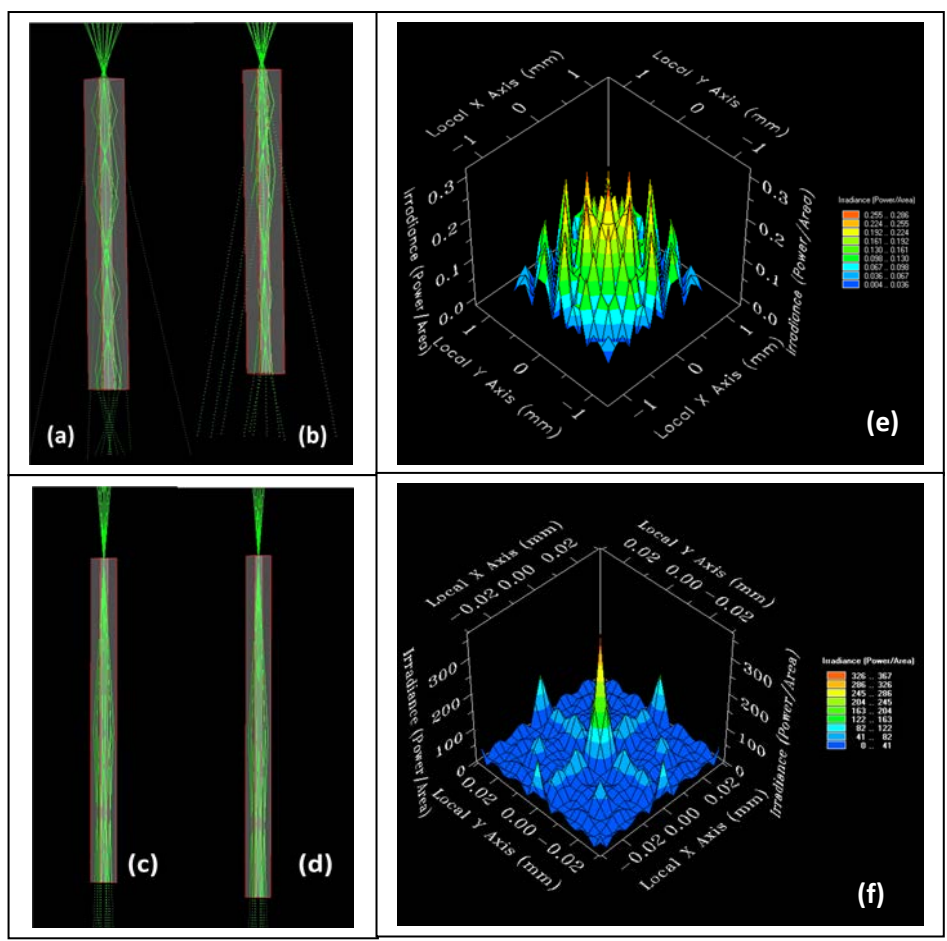

**Fig. 5: Ray-tracing simulation using FRED for two liquid waveguide structures looking from the top down. In both cases, core (liquid dye solution) indicated with the lightly shaded region which is embedded in the darker cladding region.** a. for multimode at liquid-air **interface with 110 micron width (**Waveguide II); b. for multimode at liquid-PDMS interface (Waveguide II); c. for multimode at liquid-air interface 14 micron width (Waveguide II) and d. for multimode at liquid-PDMS interface (Waveguide II) ; e. Mode field distribution in case of liquid air interface for waveguide I; f. Mode field distribution in case of liquid air interface for waveguide II

## Characterization

For any waveguide structure, there is a range of ray angle that will fulfill the total internal reflection condition based on relative refractive index difference between the core and clad region. In this case, dye solutions with different concentration act as the core medium and PDMS/air act as clad. The number of TIR for light is inversely proportional to the diameter or cross-section of microchannel. Ray tracing simulation platform (FRED) is used to understand the propagation of fluorescence light 532 nm through dye filled microstructure. Optical losses at the liquid-air interface and liquid-PDMS interface in case of multimode and single mode microstructure respectively is obtained as shown in Fig. 5. To illustrate this, Fig. 5 shows a ray-trace simulation of a liquid core waveguides. Gaussian beam from a coherent laser source is coupled at the one end of waveguide with the help of 10X objective lens for both structures. The laser light source is illuminated at the normal incidence of the waveguide. Dye solution is filled inside the microstructure. Above simulation has been applied by considering the liquid dye with R.I. of 1.4030 as the core medium embedded inside PDMS with R.I. of 1.40 and air with R.I. of 1 as the substrate. Outside the core, lower clad being PDMS (1.40) and upper clad being (Air =1), lower index region is formed.

The result obtained for different cases, shows that light can be coupled inside the microstructure filled with 1mM concentrated dye solution and confirms its waveguide nature. It also clears from this study that optical losses at liquid-air interface is comparatively less than that of liquid-PDMS interface irrespective of the dimensions of waveguide. However, dimensions of waveguide affect the total internal reflection per unit length. It is observed that waveguide structure with smaller diameter is more suitable to act as liquid mode guiding structure leading to increased probability of guiding more number of photons to reach the output end.

These results confirm that laser light can propagate through 2D liquid core waveguide structure by satisfying condition of total internal reflection over the interface of liquid core and PDMS/Air clad. By above observations, it becomes clear that many complications and challenges can be easily overcome for propagating index guided mode when air is used as a cladding material.

In this experiment, we have filled the dye solution mix of ethanol and benzyl alcohol into two microchannels (15 mm length each), with 110 micron and 14 micron width respectively, on PDMS chip. The end fire coupling method is used for optical characterization of the developed liquid waveguide structures. The schematic of characterization set up is as shown in Fig. 6 above. Here, the light from Nd:YAG laser is end coupled into waveguide I and waveguide II by using objective lens and assembly of optics is also shown in Fig. 6. The roughness of PDMS wall for 2D microchannel for both waveguide I and II were approximately limited to 1 micrometer due to the better quality of direct writing of femtosecond laser. To characterize the chip, we have used a micro syringe to insert the liquid dyes into the microchannels as the core medium. The required liquid dyes for core medium are obtained by using ethanol + benzyl alcohol as the host solution with two different solutes Rh-6G and Rh-101 to form two different dyes. Respective mixtures of these two solutes in varying concentrations act as liquid cores within the two microstructures.

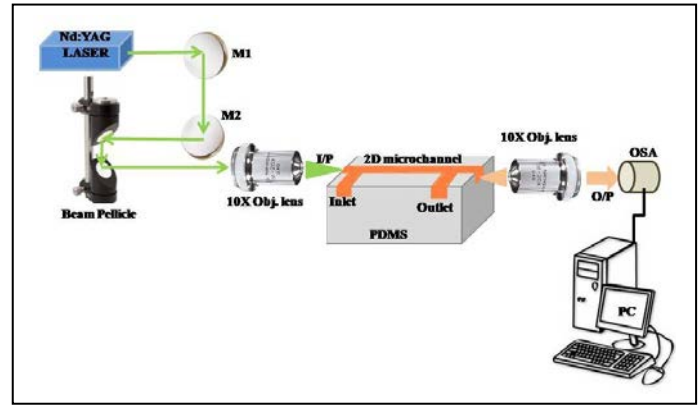

Fig. 6: Characterization setup for liquid core /Air Cladding waveguiding

As the absorption spectra of Rh-6G and Rh-101 lies in visible wavelength therefore we have selected the Nd:YAG laser with 4 mW power and 7 nsec pulse duration with rep rate 10 Hz as the pump source. This Nd: YAG laser is used to excite the fluorescent dye molecules dissolved in the liquid core. The source is aligned to beam iris and 10X objective lens. Across the objective lens beam spot size is reduced to~100 µm for waveguide II structure and 10 micron for waveguide I structure. As the light and liquid are pumped simultaneously to the microchannel, due to high refractive index difference between liquid core and air, the fluorescence light is guided and captured at the other end of microchannel. The outlet end is connected to optical spectrometer. Fluorescence spectrums are measured by changing the laser power and concentration of dyes.

**Model cross-sectional analysis for waveguide structures:** In these two structures as shown in Fig. 3 and 7, first one is multimode waveguide II structure that allows multimodal tuning of waveguides from liquid core and other one waveguide 1 support few modes propagation.

To separate the fluorescence signal and excitation light, we need 'Spectroscopic analysis and it is quite a difficult job to separate these two outputs over the output end of channel. The intensity profile for fluorescent light generated and propagated through the developed liquid waveguide structures have been measured using 'near-field intensity profile measurement' experimental set up as shown in Fig. 6 above.

The output profiles for both waveguide structures have been captured using CCD equipped with band-pass light filter for pump light (λ=532 nm). Intensity at the output end of liquid waveguide structure and corresponding intensity profile is shown in figure 7. Profile measurements make it clear that the fabricated microstructures are supporting the index guided modes for the propagation and can be used as a waveguide like structure for various applications. The small beam size (~100 µm) of the input beam, relative to that of the liquid core (100 µm), helps in reducing the coupling losses of pump light at the cross-section of the microchannel. Increment in the coupling and propagation losses are due to the increasing effects of the scattering and diffraction of the visible light through the PDMS channel walls (i.e., air/dye solution/PDMS interfaces at the front and the end) with a normal incident angle.

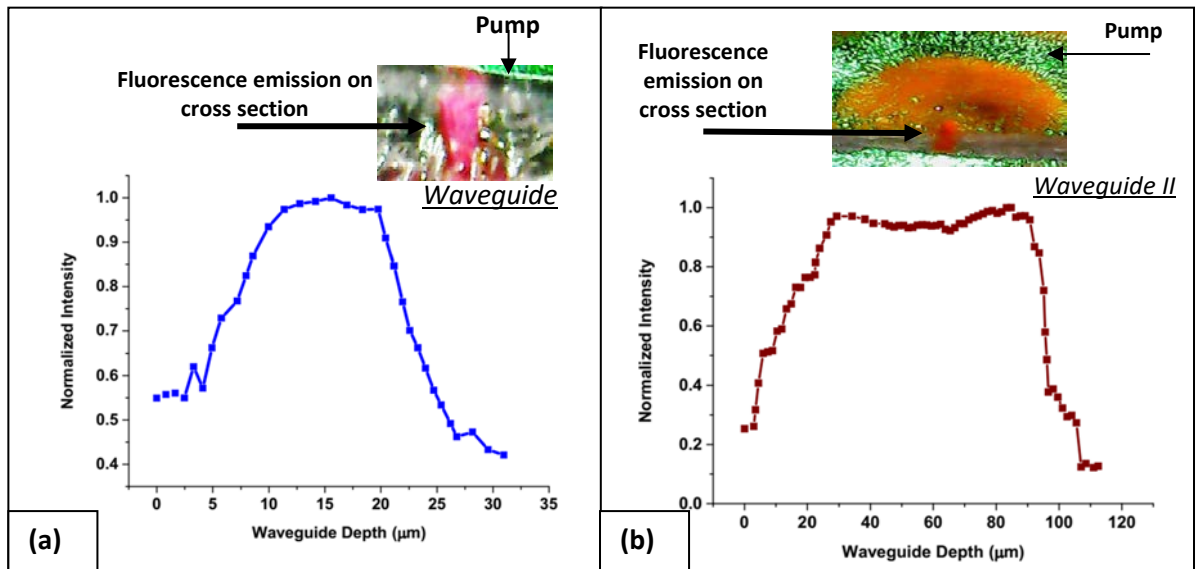

Fig. 7: Intensity distribution for light propagating through: (a) Waveguide I and (b) multimode Waveguide II liquid core/air clad cross section

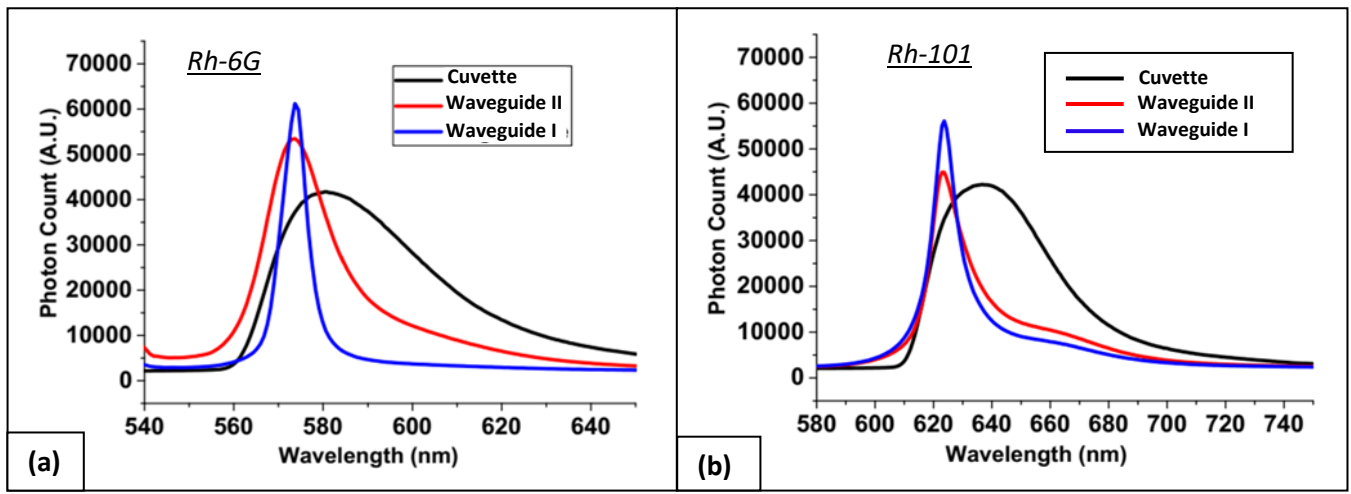

Fig. 8: Comparative studies of emission spectra for Waveguide I, Waveguide II structure and cuvette for (a)Rh-6G and (b) Rh-101 dye solution

## Results and discussion

In order to confirm waveguide nature of dye filled 2-D microstructures, we have studied the fluorescence spectroscopy for 3mM concentration of dye (Rh-6G) as a liquid medium in three different configurations i.e.) quartz Cuvette, b) waveguide II structure and c) waveguide I structure.

The fluorescence emission spectra are collected for three different structures in order to obtain the effect of microstructure dimensions on the emission output. It is observed that emission spectral peak wavelength is changed by 15 nm from microstructures to cuvette filled with same dye solution Rh-101 and pumped to a uniform Nd: YAG laser at 4 mW power as shown in Fig. 8.b. Similar shift has been observed in case of Rh-6G which shown in Fig. 8.a. Increase in output photon density confirms the coupling of FL inside the waveguide structure. It is also clear from the above fig that FWHM of FL spectra gets narrower from Cuvette to waveguide structure I. The spectral narrowing effect is observed due to the Fabry-Perot resonator formed by dye solution filled liquid waveguide and solvent-air interfaces. This result confirms that fluorescence light generated by dye solutions gets coupled through microchannel and forms Fabry-Perot type oscillations which lead us to the conclusion that 2D structure fabricated on the surface of PDMS functions as a liquid core/air cladding waveguide structure. In addition, consideration of these two waveguides and quartz cuvette confirms that dynamics of fluorescence spectra also changes. The intensity, lasing peak and line width change according to dimensions of individual structure. Same results are observed for Rh-101 dye solution. FWHM of fluorescence signal of quartz Cuvette is observed 48.8 nm and peak wavelength at 637.59 nm.

In multimode waveguide II for Rh-101 dye, line width achieved is 13.53 nm, peak wavelength is 624.10 nm and that for waveguide I structure line width is 6.94 nm and peak wavelength is 623.75nm.

In case of Rh-6Gdye solution, FWHM for Cuvette is 42.89 nm and peak wavelength is 580.90 nm. For multimode waveguide II structure is 14.52 nm, peak wavelength is 573 nm and for waveguide I structure linewidth reduces to 5.34 nm, peak wavelength is shifted at 573.70 nm. Through a comparison study, it has been observed that peak wavelength in multimode waveguide I and II structure is quite less (blue shifted) compare to Cuvette output. In the present study, we can see for quartz Cuvette the output florescence spectrum has a large bandwidth. Due to the small dimensions of microchannels, the obtained graph clearly indicates that the linewidth of waveguide structure II is less than that of Cuvette and waveguide structure I have even lower line width compared to the structure II.

**Effect of power for higher concentration regime**

For characterization of these FS written microchannels in terms of multimode waveguide microstructures I and II, we have studied the effect of pump power for Rh-6G and Rh-101 dye solutions. It has been observed that with variation in the pump power, a significant tunability has been observed in fluorescence spectra. All these measurements have been observed at the room temperature. Fig.9 illustrates the measured emission spectra with Rh-6G for 10 mM in both liquid core/air waveguide structure I and II. Here we have varied the input power in a range of 4 - 12 mW for both cases and observed that for lower concentrations, insignificant change was observed in the fluorescence peak wavelength in correspondence to the variation in incident laser power but for 10 mM, peak wavelength shift has been observed as power is varied. A florescence peak wavelength count emerges as optical pumping power density is increased. The absorption of incident laser beam is responsible for change in refractive index gradient of dye solution of the order of $10^{-3}$ to $10^{-4}$ due to optically heated thermal lensing effect [11]. Also, incident pulsed high-power laser beam generates acoustic pressure waves inside the dye filled liquid waveguide structure which induce the variation in the refractive index of medium [11, 12].

In this way, incident laser power plays significant role in the shift of florescent peak wavelength and output spectrum which is reflected in the experimental results as shown in the Fig. 9 respectively. In low concentration regime, isolated dye molecules are present but as we increase the concentration of dye, the spacing between dye molecules decreases and aggregates are formed.

Thus, peak wavelength variation can be seen in very high concentration regime. The other phenomenon which contributes to the modified output spectra of dye is 'self-absorption' due to higher concentrations. As the molecular dimmer are formed at high concentration, it explains the appearance of a second shift in measured fluorescence spectroscopy such that red shift is observed for 10 mM dye concentration by varying the power from 4 mW to 12 mW. From Fig. 9, we can clearly observe the peak wavelength for multimode waveguide structure II for Rh6G solution was achieved at 579.8 nm at 4mW pump power. As the power increases to 6mW, peak wavelength is shifted at 581.42 nm. By varying to higher power, red shifted peak wavelength is reached up to 583.25 nm. Same experiment has been repeated for Rh-101 dye solution. We took 10 mM solution and measured the fluorescence spectra for multimode waveguide structure II at 2 mW, the peak wavelength is captured at 626.48 nm. The amount of light guided inside the multimode mode waveguides I and II are strongly dependent on the refractive index difference between $n_{core}$ and $n_{clad}$ as:

$$\Delta n = n_{dye} - n_{air}$$

The Rh-6G and Rh-101 are dissolved in mixture of ethanol +benzyl alcohol as a host solution.

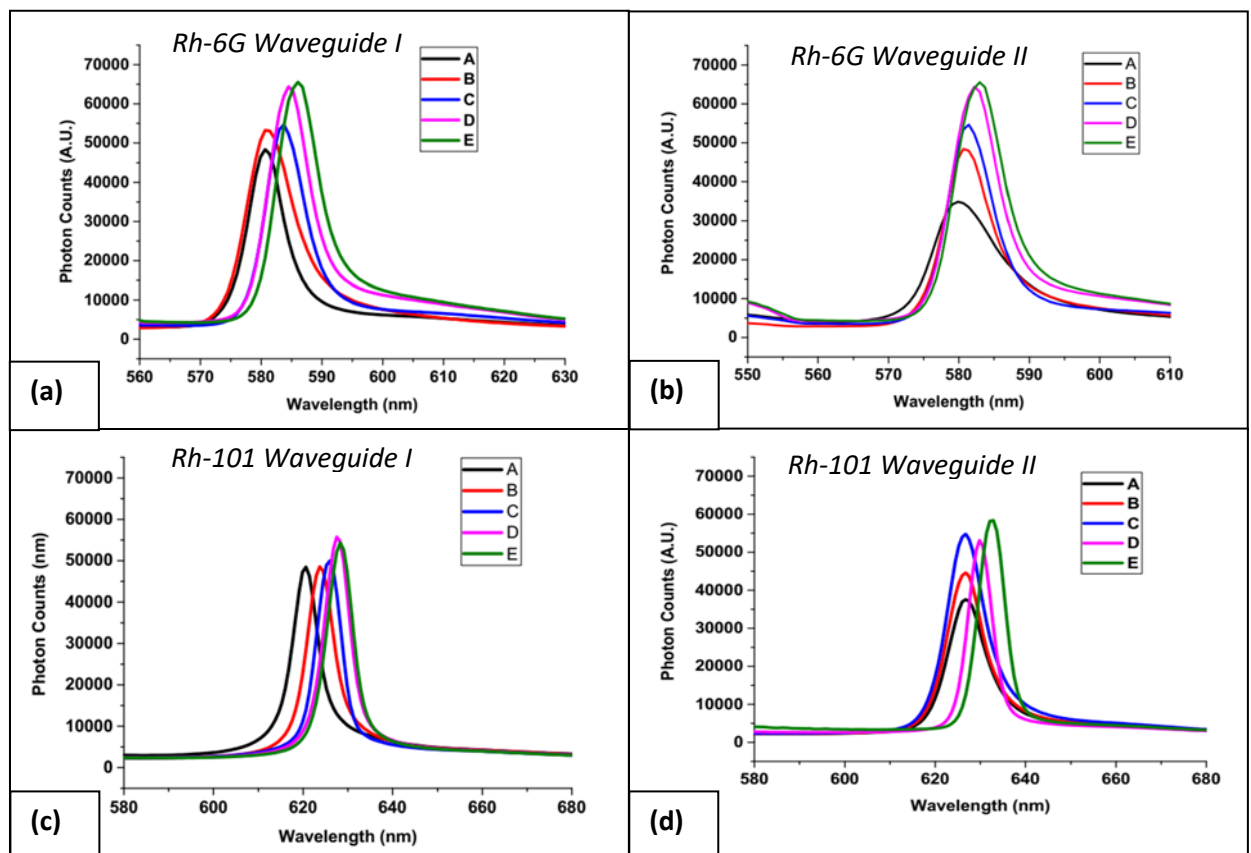

**Fig. 9:** Study of pump power-based fluorescence emission spectra for (a) Rh6G in waveguide structure I (b) Rh6G in multimode waveguide structure II (c) Rh101 in waveguide structure I and (d) Rh101 in waveguide structure II
(A=4mW, B=6 mW, C=8mW, D=10mW E=12mw)

The experiment is repeated for waveguide structure I for same solution. Light is coupled from the output end of waveguide. Light after being guided inside the waveguide structure I is observed at the cross section of the waveguide and we can observe in the graph that the peak wavelength and line width for same power and concentration changes. There is slight change in Peak wavelength but line width drastically changes in the waveguide I compared to the waveguide II.

For waveguide structure I, the red shift in to fluorescence emission peak for both dyes caused by the variation pump power from 4 mW to 12 mW with step size of 2 mW. The corresponding tunability achieved is in the range of 579.87-583.25 nm and average line width is 6.8 nm in case of Rh-6G. For the waveguide structure I, in case of Rh-101 based active solution, tunability achieved is 7 nm and observed average line width is 6 nm. For Rh-101 dye in waveguide structure I tunability achieved is in the range of 620.49-628.44 nm. In the case of waveguide structure II, for Rh-6G, red shift in peak wavelength has been observed. Tunability of peak wavelength being 4 nm and average line width being 10 nm. In the case of Rh-101, the tunability of 6 nm is achieved. For multimode waveguide structure II in case of Rh-101, spectral tunability is achieved in the range of 626.48 - 632.50 nm and average line width is 10 nm. In case of Rh-6G, for same multimode waveguide II, spectral tunability is achieved in the range of 579.87-583.25 nm and average FWHM line width is 9.5 nm.

### Effect of concentration

The tunability in output band of liquid filled microstructures is mainly determined by selection of dye solution and its solubility limit to highly dilute systems of Rh-6G and Rh-101. In case of lower concentration regime (concentrations of 0.1 mM), component of self-absorption is quite significant which decrease the intensity of signal. In addition, at higher concentrations regime (concentrations of 10 mM), intermolecular self-quenching rapidly decreased the output intensity [11]. Particularly, in high concentration regime, the Rh-6G and Rh-101 molecules arrange themselves into H type and J type dimmers [14, 15 & 16]. This dimmer formation changes the electronic structure and as a result, the output emission spectrum is also changed. In this way, the variation in the concentration of liquid medium provides an optical flexibility for liquid waveguide structures.

The experimental observation for spectral dependency of liquid waveguide structures for varying concentration of Rh-6G and Rh-101 dye solution at fixed pump power is as shown in below Fig. (10). The detailed analysis of output spectra for Rh-6G concentrations ranging from 1 mM to 4 mM and Rh-101 concentrations ranging from 1 mM to 5 mM have been done.

It was observed that the spectral position of the propagating mode through the liquid waveguide structure shifts toward longer wavelengths by increasing the concentration of dye solution. In case of waveguide I filled Rh-6G solution, the peak wavelength shift observed from 573.16 nm to 580.67nm for 1mM to 4 mM concentration change. Along with peak wavelength, average line width shift is also observed from to 5 -6.01 nm for the same. For Rh-101 filled waveguide I , 5nm shift in peak wavelength and ± 2 nm sift in line width is observed when concentration changes from 1 mM to 5 mM respectively. As Fig. 10 shows, the wavelength of the peak maximum is red-shifted with varying concentration. The same experiments were carried out for multimode waveguide structure II for both dye solutions. Similarly, spectral study for different concentrations in multimode structure II for Rh-6G dye, 8 nm red shift in peak wavelength and 1.5 nm shift in linewidth have been observed while 5 nm peak wavelength red shift with ± 2 nm linewidth shift has been observed for Rh-101 respectively. Here, the peaks occurred at different wavelengths as per the changing concentration of liquid medium. Red shift in the output spectra is observed when concentration is increased from 1 mM to 4 mM. The apparent red shift in the emitted intensity signal is due to the small Stokes shift of Rh-6G and the large spectral overlap in absorption and emission [13, 14]. Same observations have been seen for Rh-101 solution. The optimum optical absorption of pump beam, inside the dye solution filled microchannel is achieved at concentration of 1 mM.

### Photo bleaching effect in microstructure

The rate of photo bleaching primarily depends upon the type of dye, host material and their optical properties. Additionally, illumined intensity of source, wavelength of source, exposure time and temperature also affect the extent of photo bleaching [16, 17]. Photo bleaching is not a desirable phenomenon for lab-on chip based optofluidic waveguides and optofluidic lasers. It destructs the continuous output of miniaturized device and limits its usage to short time periods only. Here, we have studied the photo bleaching effect in waveguide structure I and II for both Rh-6G and Rh-101 dye mediums. This study helps us to design and improve upon the functionalities of optofluidic chips.

As a consequence of photo bleaching due to the long exposure of pump intensity to the liquid active medium, the fluorophores lose the ability to emit fluorescence in the same magnitude of intensity. The linewidth and intensity of florescence output have been significantly changed due to photo bleaching effect in the liquid waveguide. Due to diffusion dynamics in the presence of on chip reservoirs, in case of micro dye lasers, the supply of unbleached dye solution on faster time scale is not required. In the studied case, length of microchannel is 15 mm and width is 110 micron (W/L= 0.0073) for waveguide structure II and for waveguide I (W/L= 0.00093). For both the cases longitudinal coupling of light have been done in slit area. Photo bleaching time for waveguides can be converted to just a few minutes without using any costly liquid handling devices and replacement. Here, we have used the static phenomenon of liquid waveguides without using the external fluidic handling systems such as syringe pumps. The experimentally observed fluorescence dynamics is in qualitative agreement with the bleaching-diffusion dynamics [17, 18 & 19].

In Microsystems, photo bleaching creates unwanted intensity changes in the output. The quantum yield of photo bleaching and the molar extinction coefficient are inherent properties of Rhodamine-6G and Rhodamine-101. In case of static measurements inside microstructures, most affricating factors for photo bleaching inside waveguide structures filled with diluted solutions can be determined by applying Beer's law as [14]:

$$A_{out} = A_{in} \exp(-230\varepsilon Q_{ph} I_0 t_e)$$

$A_{out}$ is the amount of emitting molecules remaining after photo bleaching; $A_{in}$ is the original concentration of absorbed dye

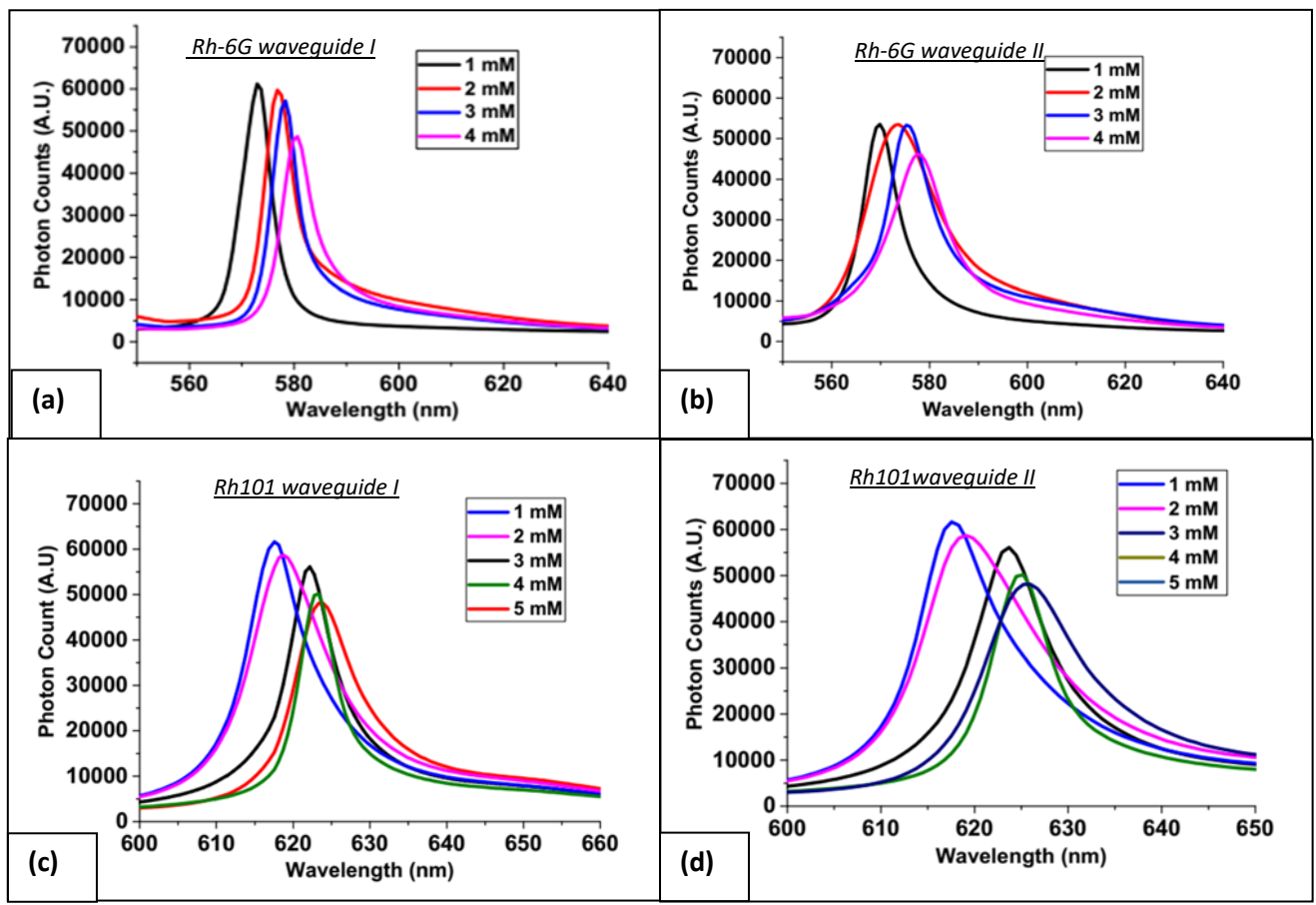

**Fig. 10:** Studies of concentration variation-based fluorescence emission spectra for multimode waveguide structures I and II for Rh6G and Rh101: (a) Rh6G filled structure I (b) Rh6G filled multimode structure II (c) Rh101 filled structure I (d) Rh101 filled multimode structure II.

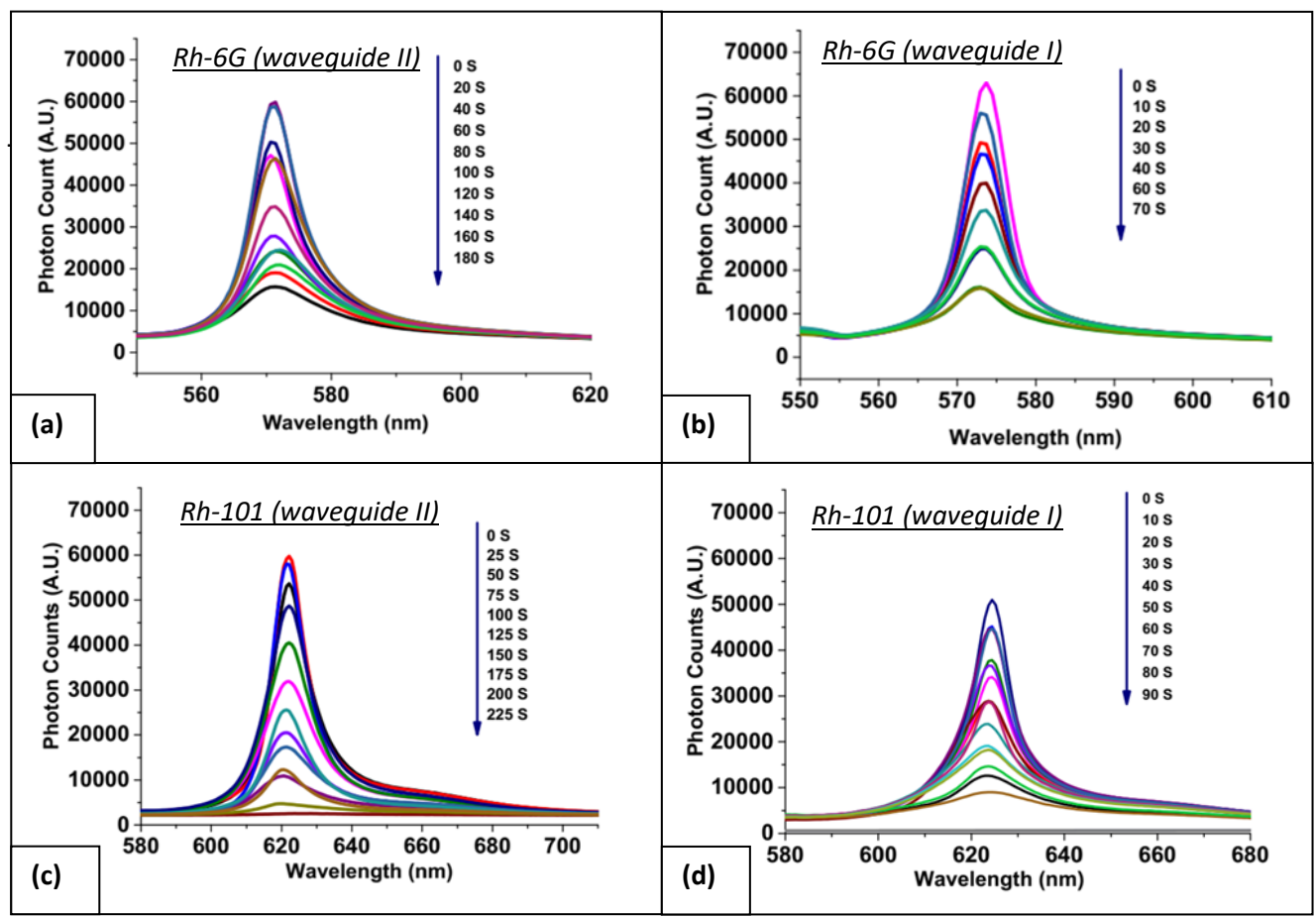

**Fig. 11:** Photobleaching studies for (a) Rh6G in multimode structure II (b) Rh6G in structure I (c) Rh101 in multimode structure II and (d) Rh101 in waveguide structure I

molecules, $I_0$ is the incident light irradiance, $Q_{ph}$ is the quantum yield of photo bleaching and $t_e$ is the exposure time. From the above equation, it is clear that quantity of photo bleached molecules inside the solution is exponentially dependent on exposure time and pump intensity. Therefore, even a small increase in time or light intensity results in a substantial increase in the amount of photo bleaching. Our experimental results reveal that these optofluidic waveguides can be operated over a few minutes without needing a flow of fresh dye solution as shown in Fig. (11). In case of Rh-6G solution, photo bleaching time is observed to be 70 sec for waveguide I and 180 sec for multimode waveguide structure II, while in case of Rh-101, photo bleaching time is observed to be 90 sec and 225 sec for multimode waveguide structure I and II respectively.

This experiment confirms that decay time of Rh-101 is slightly greater than that of Rh-6G. This observed behavior is justified by previous publication [17, 20]. The photo bleaching time can further be improved by a factor of 3 to 4 times by adding reservoirs on chip. Also, by converting the fabricated 2D structures into a 3D chip and using different pumping scheme, the developed liquid waveguide structure can be used in established optofluidic devices with enough output which would be sufficient and even more than is required to do the lab-on-chip experiments

## Conclusion:

In conclusion, we have demonstrated a novel femtosecond fabricated liquid-core/air-clad waveguide microstructures on a PDMS microchip. We have studied the role of concentration, photo-bleaching and incident power on the output of waveguides in detail. This work gives a very good understanding towards the interaction of light and fluid in micro dimension. Tunability in the form of intensity, wavelength and linewidth has been successfully obtained. The characteristic of these waveguide sources can be easily controlled and modulated by adjusting the fluid properties of the core medium. After converting these 2D chips into 3D chips and adding some optical component to the same, the liquid waveguide source can be made into a tunable optofluidic laser having a coherent light source that can be integrated with multifunctional lab-on chip systems. In this way, fluorescence measurement and detection by optofluidic devices can provide a powerful platform for analysis of biological systems and aid significantly in medical diagnostics and chemical detection. This research gives a brief idea about development and maintenance of highly functional lab-on-chip waveguides which can be used out of laboratory also for many applications.

## Acknowledgement:
We acknowledge the support provided by CMTI Bangalore, India for femtosecond micromachining fabrication facility.

## Reference:

1. B. Helbo, A. Kristensen, and A. Menon, "A micro-cavity fluidic dye laser," J. Micromech. Microeng, 2003, 13(2),307–311.
2. D. Psaltis, S. R. Quake, and C. Yang, "Developing optofluidic technology through the fusion of microfluidics and optics," Nature, 2006, 442(7101), 381–386.
3. Z. Li and D. Psaltis, "Optofluidic dye lasers," Microfluid. Nanofluidics 2008, 4 (1-2), 145–158.
4. Lin Pang,* H. Matthew Chen et al., "Optofluidic devices and applications in photonics, sensing and imaging" Lab on a Chip, 2012, 12, 3543–3551.
5. D. A. Chang-Yen, R. K. Eich, and B. K. Gale, "A monolithic PDMS waveguide system fabricated using soft-lithography techniques," J. Lightwave Technol., 2005, 23(6), 2088–2093.
6. Prashanth Reddy Konari et al.,"Experimental Analysis of Laser Micromachining of Microchannels in Common Microfluidic Substrates" Micromachines, 2021, 12, 138.
7. Felix Sima, Koji Sugioka et al.,"Three-dimensional femtosecond laser processing for lab-on-a-chip application", Nanophotonics, 2018; 7(3): 613–634.
8. Y. Yan et al., "A tunable 3D optofluidic waveguide dye laser via two centrifugal Dean flow streams", Lab on a Chip, 2011, 11, 3182.
9. Stijn Vandewiele et al., "Single-mode air-clad liquid-core waveguides on a surface energy patterned substrate", OPTICS LETTERS, 2014, Vol. 39, No. 16.
10. PengFe et al.,"A compact optofluidic cytometer with integrated liquid-core/PDMS-cladding waveguides, Lab Chip, 2012, 12, 3700–3706.
11. S.k Mishra et.al. "Measurement of Thermo Optical Coefficient for Commonly used Dye Solvents", International journal of photonics and optical technology, 2018,Vol. 4, Iss. 2, pp: 12-16.
12. Shane M. Eaton,Carmela De Marco,Rebeca Martinez-Vazquez,Roberta Ramponi,Stefano Turri,Giulio Cerullo,Roberto Osellame, "Femtosecond laser microstructuring for polymeric lab-on-chips" Journal of Biophotonics, 2012, 5(8-9).
13. Penzkofer. W. I.eupacher et al., " Fluorescence behaviour of highly concentrate rhodamine 6G solutions". journal of Luminescence, 1987,37, 61-72.
14. Florian M. Zehentbauer et al., "Fluorescence spectroscopy of Rhodamine 6G: Concentration and solvent effects", SpectrOChimica Acta Pan A: Molecular and Biomolecular SpectrOscopy, 2014, 121() 147-151.
15. K. Noack, J. Kiefer, A.I.saipem, et al., "Concentration dependent hydrogen bonding effects on the dimethyl sulfoxide vibrational structure in the presence of water, methanol and ethano"l, ChemPhysChem 2010, 11, 630-637.
16. VJ. Gavrilenko, MA Noginov, et al., "Ab initio study of optical properties of Rhodamine 6G molecular dimers", Journal of Chemical Physic, 2006s 124, 044301.
17. Morten Gersborg-Hansen et al., "Bleaching and diffusion dynamics in optofluidic dye lasers", APPLIED PHYSICS LETTERS, 2007,90, 143501.
18. Jerker Widengreny et al., "Mechanisms of photobleacing investigated by fluorescence correlation spectroscopy", Bioimaging, 1996, 4, 149–15.
19. Mingyu Chapma et. al., "Rhodamine 6G Structural Changes in Water/Ethanol Mixed Solvent", Journal of Fluorescence, 2018, 28:1431–1437.
20. Julien Laverdant et. al. , "Experimental Determination of the Fluorescence Quantum Yield of Semiconductor Nanocrystals", Materials, 2011, 4, 1182-1193.